# Novel Design of Quantum Circuits for Representation of Grayscale Images

Mayukh Sarkar

Department of Computer Science and Engineering,
Motilal Nehru National Institute of Technology Allahabad, Prayagraj, India
mayukh@mnnit.ac.in

**Abstract.** The advent of quantum computing has influenced researchers around the world to solve multitudes of computational problems with the promising technology. Feasibility of solutions for computational problems, and representation of various information, may allow quantum computing to replace classical computer in near future. One such challenge is the representation of digital images in quantum computer. Several works have been done to make it possible. One such promising technique, named Quantum Probability Image Encoding [1-2], requires minimal number of qubits, where the intensity of $n$ pixels is represented as the statevector of $\lceil \log_2 n \rceil$ qubits. Though there exist quantum circuit design techniques to obtain arbitrary statevector, they consider statevector in general Hilbert space. But for image data, considering only real vector space is sufficient, that may constraint the circuit in smaller gate set, and possibly can reduce number of gates required. In this paper, construction of such quantum circuits has been proposed.

## 1    Introduction

Quantum computing, one of the buzzwords in today's research, started from the idea of quantum mechanical model of Turing machine proposed by Paul Benioff, in 1980 [3]. In 1982, Richard Feynman, ideate the possibility of quantum computer [4], and it became a buzzword when, in 1994, Peter Shor proved its capability by proposing a quantum polynomial time algorithm of integer factorization [5]. Since then, researchers all around the world have been trying to solve multitudes of computational problems using this technology. One of the promising applications is the domain of image processing using this powerful paradigm.

To implement any image processing algorithm on a quantum computer, it is first necessary to represent the image using qubits. There have been a multitude of techniques devised for representation of an image on a quantum computer, such as Real Ket [6], Qubit Lattice [7], Entangled image [8], FRQI (Flexible representation of quantum images) [9], NEQR (Novel quantum representation) [10], NCQI (a novel quantum representation of color digital images) [11], QPIE (quantum probabilistic image encoding) [1-2] etc., and their further improvements. A good overview of major quantum image representations, along with an improvement of NCQI, named INCQI, can be found in Su et. al. [12]. Among these techniques, QPIE represents the pixels of a grayscale image as probability amplitudes of statevector of a possible quantum state. The advantage of this technique is requirement of minimal number of qubits, namely $\lceil \log_2 n \rceil$ qubits for representation of $n$ pixels. Higher the number of qubits used in a system, it becomes more costly and error-prone. Added advantages is the straightforward implementation of edge detection [1], if pixels are represented using QPIE technique.

But the techniques used in quantum circuit design for arbitrary statevector [13-14], considers general statevectors in complete Hilbert space. But image data do not require complex numbers, and hence it is sufficient to consider only real vector space. This allows us to confine the gates used in circuit within NCT (NOT, CNOT, Toffoli), and $R_y$ gates, and their controlled counterparts. It has been shown in Section 3.2 that, this technique also requires lesser number of gates than circuit obtained using generalized techniques [13-14], at least for small circuits. This paper proposes an algorithm to design quantum circuits consisting of only gates with real matrices, to represent an arbitrary real statevector, which can be utilized to represent pixels of an grayscale image on a quantum computer following QPIE technique.

Section 2 gives a brief overview of QPIE techniques. Section 3 describes the complete proposed work, in which first single-qubit and two-qubits circuits are designed as base cases for smaller statevectors, followed by general recursive algorithm to design quantum circuit for real statevectors of any dimension. Section 4 then finally concludes the work.

## 2    Background Information

Several representations of digital images have been proposed in quantum computing literature over last decade, some of which has been mentioned in Para 2 of Section 1. Among the proposed techniques, the most common ones used in practice are FRQI, NEQR, and QPIE. Yao et.al. [1], in the base paper of QPIE, have proposed a quantum image representation technique in which pixel values are encoded as probability amplitudes. In this section, an overview of QPIE technique is being provided.

In QPIE, the 2-dimensional image data is first unfolded into a vector form. If $I = (f_{ij})_{M \times L}$ be an image data, where $f_{ij}$ represents pixel value at position (i, j), M and L are number of rows and number of columns respectively, it is first unfolded into vector *v* as,

$$v = \begin{bmatrix} f_{11}, & f_{21}, & \ldots, f_{M1}, & f_{21}, & f_{22}, & \ldots, & f_{M2}, & \ldots, f_{ij}, & \ldots, f_{ML} \end{bmatrix}^T \qquad (4)$$

Then this image data is encoded into a quantum state

$$|I\rangle = \sum_{k=0}^{2^n-1} c_k \, |k\rangle \qquad (5)$$

of $n = \lceil \log_2(ML) \rceil$ qubits. $|k\rangle$ represents the computational basis encoding the position (i, j) and $c_k = \frac{f_{ij}}{\sqrt{\Sigma(f_{ij})^2}}$ represents the pixel values encoded as probability distribution satisfying $\Sigma|c_k|^2 = 1$.

Given that, $c_k$ and $\Sigma|c_k|^2$ can be calculated efficiently, the *n*-qubit state representing the image data can be created efficiently in *O(poly(n))* steps, where *poly(n)* represents some polynomial function of *n* [1]. Arbitrary state preparation techniques proposed by Grover et. al. [13] and Soklakov et.al. [14], includes unit vectors in $2^n$-dimensional Hilbert space, i.e., vectors may contain complex amplitudes. But pixel data of an image is always real. Though the generalized state preparation techniques can also prepare such states, removing the necessity of handling complex amplitudes has the ability to obtain circuits with smaller subset of gates, such as NCT, and $R_y$ gates and their controlled counterparts, which keeps the states only in real vector space. Current paper proposes exactly the same, i.e., goal of the current paper is to propose an algorithm to produce a quantum circuit producing state as in equation (5), solely for unit vectors in real vector space, such as normalized pixel data of a grayscale image.

Note that, as the pixel data is being represented as probability amplitudes of a quantum statevector, it cannot be used to store the image for further retrieval, as measuring the statevector will collapse the complete quantum state, thereby destroying the complete pixel data. This work is expected to be important in the applications requiring state preparation circuits, where an image needs to be represented using minimal number of qubits, temporarily. These qubits are then further processed via the image processing circuit, performing important image processing applications.

## 3 Proposed Work

In this section, the technique to generate a quantum circuit with *n* qubits that will produce an arbitrary unit vector in $2^n$-dimensional real vector space, is proposed. The circuit consists of only NCT, and $R_y$ gates and their controlled counterparts. To demonstrate the technique, let us first start with a 2-dimensional unit real vector.

### 3.1 Single qubit circuit generating 2-dimensional arbitrary real statevector

Let us consider an arbitrary real vector $|\psi\rangle = \begin{bmatrix} \alpha_1 \\ \alpha_2 \end{bmatrix}$ with $\alpha_1^2 + \alpha_2^2 = 1$. Thus we can readily consider $\alpha_1 = \cos\frac{\theta}{2}$ and $\alpha_2 = \sin\frac{\theta}{2}$ for certain angle $\theta$, which can be obtained as $\theta = 2\arccos(\alpha_1)$. The following circuit will generate the desired state.

$$|0\rangle \;-\!\!\boxed{R_y(\theta)}\!\!-\; |\psi\rangle$$

**Fig. 1.** Quantum circuit to generate arbitrary 2-D real statevector

### 3.2 Two qubit circuit generating 4-dimensional arbitrary real statevector

Now, let us consider an arbitrary real vector $|\psi\rangle = \begin{bmatrix} \alpha_1 \\ \alpha_2 \\ \alpha_3 \\ \alpha_4 \end{bmatrix}$ satisfying $\alpha_1^2 + \alpha_2^2 + \alpha_3^2 + \alpha_4^2 = 1$. Thus we can consider three real angles $\theta_1, \theta_2, \theta_3$ such that $\alpha_1 = \cos\frac{\theta_1}{2}$, $\alpha_2 = \sin\frac{\theta_1}{2}\cos\frac{\theta_2}{2}$, $\alpha_3 = \sin\frac{\theta_1}{2}\sin\frac{\theta_2}{2}\cos\frac{\theta_3}{2}$, and $\alpha_4 = \sin\frac{\theta_1}{2}\sin\frac{\theta_2}{2}\sin\frac{\theta_3}{2}$. This is in accordance with the spherical coordinate system.

Now, with initial state of a two-qubit quantum system being $(1, 0, 0, 0)^{\mathrm{T}}$, the circuit generating the desired quantum state can be designed as follows.

(a) Employing $R_y(\theta_1)$ on first qubit yields the state $(\cos\frac{\theta_1}{2}, \sin\frac{\theta_1}{2}, 0, 0)^{\mathrm{T}}$, following the similar logic in Section 3.1.

(b) Employing controlled-$R_y(-\theta_2)$ gate with control on first qubit and target on second qubit performs following operation.

$$\begin{bmatrix} 1 & 0 & 0 & 0 \\ 0 & \cos\left(-\frac{\theta_2}{2}\right) & 0 & -\sin\left(-\frac{\theta_2}{2}\right) \\ 0 & 0 & 1 & 0 \\ 0 & \sin\left(-\frac{\theta_2}{2}\right) & 0 & \cos\left(-\frac{\theta_2}{2}\right) \end{bmatrix}\begin{bmatrix} \cos\frac{\theta_1}{2} \\ \sin\frac{\theta_1}{2} \\ 0 \\ 0 \end{bmatrix} = \begin{bmatrix} \cos\frac{\theta_1}{2} \\ \sin\frac{\theta_1}{2}\cos\frac{\theta_2}{2} \\ 0 \\ -\sin\frac{\theta_1}{2}\sin\frac{\theta_2}{2} \end{bmatrix}$$

(c) Employing controlled-$R_y(\pi + \theta_3)$ gate with control on second qubit and target on first qubit performs following operation.

$$\begin{bmatrix} 1 & 0 & 0 & 0 \\ 0 & 1 & 0 & 0 \\ 0 & 0 & \cos\left(\frac{\pi}{2}+\frac{\theta_3}{2}\right) & -\sin\left(\frac{\pi}{2}+\frac{\theta_3}{2}\right) \\ 0 & 0 & \sin\left(\frac{\pi}{2}+\frac{\theta_3}{2}\right) & \cos\left(\frac{\pi}{2}+\frac{\theta_3}{2}\right) \end{bmatrix}\begin{bmatrix} \cos\frac{\theta_1}{2} \\ \sin\frac{\theta_1}{2}\cos\frac{\theta_2}{2} \\ 0 \\ -\sin\frac{\theta_1}{2}\sin\frac{\theta_2}{2} \end{bmatrix} = \begin{bmatrix} \cos\frac{\theta_1}{2} \\ \sin\frac{\theta_1}{2}\cos\frac{\theta_2}{2} \\ \sin\frac{\theta_1}{2}\sin\frac{\theta_2}{2}\cos\frac{\theta_3}{2} \\ \sin\frac{\theta_1}{2}\sin\frac{\theta_2}{2}\sin\frac{\theta_3}{2} \end{bmatrix}$$

The output state, as observed, matches with our desired statevector. The circuit thus demonstrated, is shown as in figure 2.

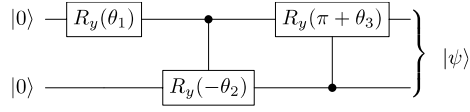

**Fig. 2.** Quantum circuit to generate arbitrary 4-D real unit vector

The generated circuit has been tested on several randomly generated 4-dimensional real array with elements in range [0, 255], using Qiskit library in Python.

As an example, when the above-mentioned procedure is employed on the pixel data [0, 128, 192, 255], the following quantum circuit, as shown in figure 3, is produced.

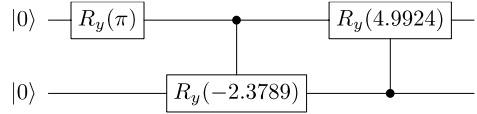

**Fig 3.** Quantum circuit to generate statevector corresponding to [0, 128, 192, 255].

Whereas, as accessed on the day of this writing, the circuit proposed by the Qiskit tutorial website [2], for the 4-pixel image with pixel values [0, 128, 192, 255], consists of 5 quantum gates.

### 3.3 Multi-qubit circuit generation for arbitrary $2^n$-dimensional real statevector.

Suppose we have been given any arbitrary grayscale image. We can readily pad the image with zeros to make number of pixels as power of 2. Let number of pixels, after padding, turns out to be $2^n$. After scaling and converting it into probabilistic amplitudes of a possible quantum system statevector, the $n$-qubit quantum circuit to generate the arbitrary $2^n$-dimensional statevector, can be obtained as follows. Generation of a 3-qubit quantum circuit for 8-dimensional statevector, is being shown as example along with each step.

(a) Obtain spherical angles from the statevector. With $2^n$-dimensional statevector, we will obtain $(2^n - 1)$ angles. As an example, for an 8-dimensional statevector $[c_0, c_1, c_2, c_3, c_4, c_5, c_6, c_7]^T$, we can obtain 7 spherical angles $[\alpha_0, \alpha_1, \alpha_2, \alpha_3, \alpha_4, \alpha_5, \alpha_6]$ such that, the statevector can be represented as $[\cos\frac{\alpha_0}{2}, \ sin\frac{\alpha_0}{2}cos\frac{\alpha_1}{2}, \ sin\frac{\alpha_0}{2}sin\frac{\alpha_1}{2}cos\frac{\alpha_2}{2}, \ sin\frac{\alpha_0}{2}sin\frac{\alpha_1}{2}$
$sin\frac{\alpha_2}{2}cos\frac{\alpha_3}{2}, \ sin\frac{\alpha_0}{2}sin\frac{\alpha_1}{2}sin\frac{\alpha_2}{2}sin\frac{\alpha_3}{2}cos\frac{\alpha_4}{2}, \ sin\frac{\alpha_0}{2}sin\frac{\alpha_1}{2}sin\frac{\alpha_2}{2}sin\frac{\alpha_3}{2}$

$sin\frac{\alpha_4}{2}cos\frac{\alpha_5}{2}$, $\sin\frac{\alpha_0}{2}\sin\frac{\alpha_1}{2}sin\frac{\alpha_2}{2}sin\frac{\alpha_3}{2}sin\frac{\alpha_4}{2}sin\frac{\alpha_5}{2}cos\frac{\alpha_6}{2}$, $\sin\frac{\alpha_0}{2}\sin\frac{\alpha_1}{2}$ $sin\frac{\alpha_2}{2}sin\frac{\alpha_3}{2}sin\frac{\alpha_4}{2}sin\frac{\alpha_5}{2}sin\frac{\alpha_6}{2}]$.

(b) If $n = 1$ or $2$, employ design techniques mentioned in Sections 3.1 and 3.2 respectively. Otherwise, design an (n-1)-qubit arbitrary statevector generator circuit, recursively, employing the first (n-1) qubits on the system. This will involve first $(2^{n-1} - 1)$ spherical angles, and will build up first $(2^{n-1} - 1)$ entries of the statevector completely, and $2^{n-1}$ th entry partially. As an example, for the 3-qubit system, employ the design of Section 3.2 with first $(2^{n-1} - 1) = 3$ angles, as shown in figure 4. The output of the partial circuit is $[\cos\frac{\alpha_0}{2}$, $\sin\frac{\alpha_0}{2}\cos\frac{\alpha_1}{2}$, $\sin\frac{\alpha_0}{2}\sin\frac{\alpha_1}{2}\cos\frac{\alpha_2}{2}$, $\sin\frac{\alpha_0}{2}\sin\frac{\alpha_1}{2}\sin\frac{\alpha_2}{2}, 0, 0, 0, 0]$. Observe that $c_3$ has been created partially.

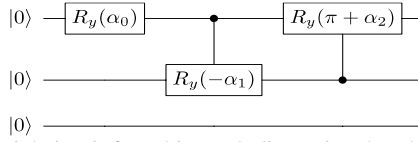

**Fig 4**. Partial circuit for arbitrary 8-dimensional real statevector

(c) Employ an (n-1)-qubit controlled $R_y(\alpha_3)$ gate, with control on first (n-1) qubits and target on last qubit. This will employ $R_y(\alpha_3)$ on entries $0\underbrace{111\ldots111}_{n-1}$ ($2^{n-1}$ th entry) and $1\underbrace{111\ldots111}_{n-1}$ ($2^n$ th entry). The circuit in figure 5 has the output as $[\cos\frac{\alpha_0}{2}$, $\sin\frac{\alpha_0}{2}\cos\frac{\alpha_1}{2}$, $\sin\frac{\alpha_0}{2}\sin\frac{\alpha_1}{2}cos\frac{\alpha_2}{2}$, $\sin\frac{\alpha_0}{2}\sin\frac{\alpha_1}{2}sin\frac{\alpha_2}{2}\cos\frac{\alpha_3}{2}$, $0, 0, 0, \sin\frac{\alpha_0}{2}\sin\frac{\alpha_1}{2}sin\frac{\alpha_2}{2}sin\frac{\alpha_3}{2}]$.

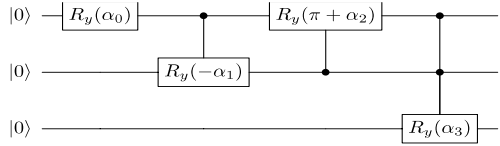

**Fig 5**. Partial circuit for arbitrary 8-dimensional real statevector

(d) Employ (n-1) CNOT gates, one by one, on each of the first (n-1) qubits. Each of these CNOT gates has control on the last qubit. These (n-1) gates will take the entry at $\underbrace{1111\ldots111}_{n}$ (last entry) to $1\underbrace{000\ldots000}_{n-1}$ ($2^{n-1}+1$ th entry). The circuit in figure 6 has the output $[\cos\frac{\alpha_0}{2}$, $\sin\frac{\alpha_0}{2}\cos\frac{\alpha_1}{2}$, $\sin\frac{\alpha_0}{2}\sin\frac{\alpha_1}{2}cos\frac{\alpha_2}{2}$, $\sin\frac{\alpha_0}{2}\sin\frac{\alpha_1}{2}sin\frac{\alpha_2}{2}\cos\frac{\alpha_3}{2}$, $\sin\frac{\alpha_0}{2}\sin\frac{\alpha_1}{2}sin\frac{\alpha_2}{2}sin\frac{\alpha_3}{2}, 0$, $0, 0]$.

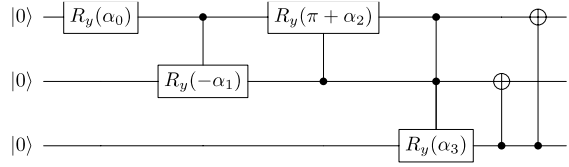

**Fig 6.** Partial circuit for arbitrary 8-dimensional real statevector

(e) Employ another (n-1)-qubit arbitrary statevector generator circuit with last $(2^{n-1} - 1)$ angles, recursively, on first (n-1) qubits. Each gate in this sub-circuit must have additional control from last qubit. The final 3-qubit circuit is shown in figure 7. It has 2-qubit arbitrary statevector generator circuit of figure 2 with angles $[\alpha_4, \alpha_5, \alpha_6]$, employed after the partial circuit of figure 6, each having additional control from last qubit.

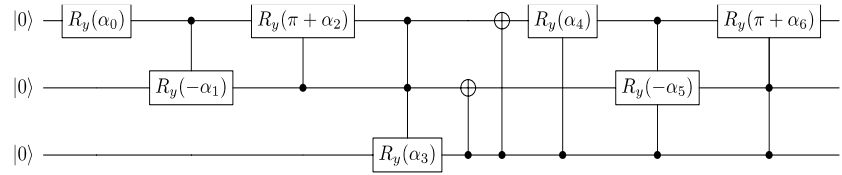

**Fig 7.** Complete quantum circuit to generate arbitrary 8-dimensional real statevector

The circuit of figure 7 eventually has the desired output statevector. The circuit thus designed, has also been verified successfully with Qiskit library in Python, on several randomly generated 8-dimensional arrays with values in range [0, 255].

## 4    Conclusion

In this work, a novel technique to design a quantum circuit in order to create an arbitrary real statevector has been proposed. Designed circuit is composed solely of NCT, and $R_y$ gate and its controlled counterparts. The circuit needs $O(poly(n))$ gates, where *n* is number of pixels in the image. Though the designed circuit may not have minimal number of gates performing the task, number of gates have been found to be lower than circuits designed using generalized techniques [13-14], at least for small circuits. This work is expected to be important in the applications requiring state preparation circuits, where an image needs to be represented using minimal number of qubits, temporarily. The qubits representing pixel data can then be further processed via an image processing circuit for important and interesting applications.

**Acknowledgements.** This paper would not have been possible without the financial support of MeitY Quantum Computing Applications Lab (QCAL). The author is also thankful to Amazon AWS for providing the great environment for Quantum Computing research, that was needed to perform the experiments using the Qiskit provider for Amazon Braket.